\documentclass[aps,pre,twocolumn,showpacs,superscriptaddress,preprintnumbers,amsmath,amssymb]{revtex4}
\usepackage{amsmath}
\usepackage{graphicx}% Include figure files
\usepackage{dcolumn}% Align table columns on decimal point
\usepackage{bm}% bold math
\usepackage{float}

\begin{document}

\title{Emergence of MHD structures in a collisionless PIC simulation plasma}

\author{M. E. Dieckmann}
\affiliation{Department of Science and Technology, Link\"oping University, SE-60174 Norrk\"oping, Sweden}
\author{D. Folini}\affiliation{\'Ecole Normale Sup\'erieure, Lyon, CRAL, UMR CNRS 5574, Universit\'e de Lyon, France}
\author{R. Walder}\affiliation{\'Ecole Normale Sup\'erieure, Lyon, CRAL, UMR CNRS 5574, Universit\'e de Lyon, France}
\author{L. Romagnani}\affiliation{\'Ecole Polytechnique, CNRS, LULI, F-91128 Palaiseau, France}
\author{E. d'Humieres}\affiliation{Univ Bordeaux, IMB, UMR 5251, F-33405 Talence, France}
\author{A. Bret}\affiliation{ETSI Industriales, Universidad de Castilla-La Mancha, 13071 Ciudad Real and Instituto de Investigaciones Energ\'{e}ticas y Aplicaciones Industriales, Campus Universitario de Ciudad Real, 13071 Ciudad Real, Spain}
\author{T. Karlsson}\affiliation{KTH Royal Inst Technol, Sch Elect Engn, Space \& Plasma Phys, Stockholm, Sweden}
\author{A. Ynnerman} 
\affiliation{Department of Science and Technology, Link\"oping University, SE-60174 Norrk\"oping, Sweden}

\date{\today}% It is always \today, today,

% PACS, the Physics and Astronomy Classification Scheme. %%

\begin{abstract}
The expansion of a dense plasma into a dilute plasma across an initially uniform perpendicular magnetic field is followed with a one-dimensional particle-in-cell (PIC) simulation over MHD time scales. The dense plasma expands in the form of a fast rarefaction wave. The accelerated dilute plasma becomes separated from the dense plasma by a tangential discontinuity at its back. A fast magnetosonic shock with the Mach number 1.5 forms at its front. Our simulation demonstrates how wave dispersion widens the shock transition layer into a train of nonlinear fast magnetosonic waves.
\end{abstract}

\maketitle
A thermal pressure gradient in a magnetized plasma accelerates the dense plasma towards the dilute one and a rarefaction wave develops. The collision of the expanding plasma with the dilute plasma triggers shocks if the collision speed exceeds the phase velocity of the fastest ion wave. In the rest frame of the shock, the fast-moving upstream plasma is slowed down, compressed and heated as it crosses the shock and moves downstream. This net flux adds material to the downstream plasma, which lets the shock expand into the upstream direction. 

Shocks have been widely examined due to their key role in regulating the transfer of mass, momentum and energy in plasma. They are most easily described in a one-dimensional geometry. Shock tube experiments \cite{Dolder60,Borisov71}, which enforce such a geometry, examined shocks in partially magnetized and collisional plasma. Particle collisions equilibrate the plasma and macroscopic quantities like flow speed and temperature are uniquely defined. The time-evolution of these quantities is described well by the equations of single-fluid magnetohydrodynamics (MHD) if collisions are frequent enough to establish a thermal equilibrium between electrons and ions on the time scales of interest. Numerical shock tube experiments investigating the thermal expansion of plasma have also been performed in order to test single-fluid MHD codes, since the important MHD shocks emerge under such conditions \cite{Brio88,Falle98}. 

However, not all plasma shocks are collisional. The mean free path of the particles in the plasma, in which the Earth's bow shock \cite{Balogh05} is immersed, is large compared to the thickness of its transition layer and it is sustained by electromagnetic fields. Collisionless plasmas support energetic structures that are not captured by a single-fluid MHD theory and that can play a vital role in the thermalization of plasma. Examples are magnetosonic solitons \cite{Stasiewicz03,Guerolt17} and the beams of shock-reflected particles ahead of the bow shock \cite{Eastwood05}, which enforce a non-stationarity of the shock \cite{Chapman05,Burgess07,Marcowith16,Sundberg17}. Single-fluid MHD simulations are nevertheless used to solve problems in collisionless plasma based on the argument that they can describe the plasma dynamics on a large enough scale. 

Here we examine with the particle-in-cell (PIC) code EPOCH \cite{Arber15} the relaxation of a thermal pressure gradient in the presence of a perpendicular magnetic field. We thus perform a numerical shock tube experiment with collisionless plasma to test the hypothesis that the plasma evolution will resemble its equivalent in MHD. The plasma parameters are within reach for laser-plasma experiments and our results can thus be tested experimentally. The expansion speed of our blast shell remains below that considered in Ref. \cite{Guerolt17} and no shock reformation takes place. We use the same setup as in Ref. \cite{Dieckmann16}, where we investigated the initial evolution of the expanding plasma and observed a lower-hybrid wave shock at the front of the expanding plasma, which has no counterpart in a single-fluid theory. Here we show that this kinetic shock is transient and that the plasma dynamics is eventually regulated by structures that exist also in the single-fluid MHD model \cite{Myong97}.

Our simulation setup is as follows: we resolve one spatial dimension and three particle velocity components. Open boundary conditions are used for the fields and reflecting boundary conditions are used for the computational particles (CPs). The simulation box is large enough to separate effects introduced by the boundaries from the area of interest. The length $L_0$ = 0.75 m of the simulation box is subdivided into evenly spaced grid cells with the length $\Delta_x = 5\mu m$. The particle dynamics is determined in PIC simulations exclusively by the charge-to-mass ratio. We consider here the fully ionized nitrogen that is frequently used in laser plasma experiments. The plasma in the interval $0 < \hat{x} < 2L_0/3$ consists of ions with the number density $n_0$ and electrons with the number density $7n_0 = 2.75 \times 10^{20} \textrm{m}^{-3}$. The electron temperature is $T_e = 2.32 \times 10^7$ K and the ion temperature $T_i = T_e /12.5$. We refer to this dilute plasma as ambient plasma. A denser plasma is located in the interval $-L_0/3 \le \hat{x} \le 0$. It consists of ions with the density $10 n_0$ and the temperature $T_i$. Its electrons have the density $70n_0$ and the temperature $3T_e$. All species are initially at rest. A spatially uniform background magnetic field with the strength $B_0 = 0.85$ T is aligned with z. We represent the electrons and ions of the ambient plasma by $3\times 10^7$ CPs each while the electrons and ions of the dense plasma are each resolved by $4.5 \times 10^7$ CPs.

The values for the parameters of the ambient plasma are listed in Table \ref{table1} ($e$, $\mu_0$, $m_e$, $m_i$, $k_B$, $\gamma_e=5/3$, $\gamma_i=3$: elementary charge, permeability, electron mass, ion mass, Boltzmann constant and adiabatic constants for electrons and ions). These parameters are the electron plasma frequency $\omega_{pe}$ and gyro-frequency $\omega_{ce}$, the electron thermal speed $v_{the}$ and thermal gyroradius $r_{ge}$, the ion plasma frequency $\omega_{pi}$ and gyro-frequency $\omega_{ci}$, the lower-hybrid frequency $\omega_{lh}$, the ion acoustic speed $c_s$, the Alfv\'en speed $v_A$ and the fast magnetosonic speed $V_{fms}$.
\begin{table}
\begin{tabular}{ll}
Parameter & Numerical value \\
$\omega_{pe}={(n_e e^2/\epsilon_0 m_e)}^{1/2}$ & $9.35 \times 10^{11}\mathrm{s}^{-1}$\\
$\omega_{ce}=eB_0/m_e$ & $1.5 \times 10^{11} \mathrm{s}^{-1}$\\
$v_{the}={(k_B T_e / m_e)}^{1/2}$ & $1.87 \times 10^7 \mathrm{ms}^{-1}$\\
$r_{ge}=v_{the}/\omega_{ce}$ & $1.25 \times 10^{-4}\mathrm{m}$\\
$\omega_{pi}={(Z^2 n_i e^2/\epsilon_0 m_i)}^{1/2}$ & $1.54 \times 10^{10}\mathrm{s}^{-1}$\\
$\omega_{ci}=ZeB_0/m_i$ & $4.07 \times 10^7s^{-1}$\\
$\omega_{lh}= {({(\omega_{ce}\omega_{ci})}^{-1}+\omega_{pi}^{-2})}^{-1/2}$ & $2.46 \times 10^{9}\mathrm{s}^{-1}$\\
$c_s={((\gamma_eZk_BT_e + \gamma_ik_BT_i)/m_i)}^{1/2}$ & $4.03 \times 10^5 \mathrm{m}\mathrm{s}^{-1}$ \\
$v_A = B_0/{(\mu_0n_0m_i)}^{1/2}$ & $7.9 \times  10^5\mathrm{m}\mathrm{s}^{-1}$ \\
$V_{fms}={(c_s^2 + v_A^2)}^{1/2}$ & $8.7 \times 10^{5}\mathrm{m}\mathrm{s}^{-1}$\\
\end{tabular}\caption{The plasma parameters in our simulation.}\label{table1}
\end{table}

%A spatially uniform background magnetic field with the strength $B_0 = 0.85$ T is aligned with z. The electron thermal speed $v_{the}={(k_B T_e / m_e)}^{0.5}$ and gyro-frequency $\omega_{ce} = eB_0/m_e$ ($e$, $m_e$, $k_B$: elementary charge, electron rest mass and Boltzmann constant) determine their thermal gyroradius $r_{ge}=v_{the}/\omega_{ce}$ in the ambient plasma, which will become the upstream plasma. The gyro-frequency, the plasma frequency and the lower-hybrid frequency of ions with $Z=7$ and the mass $m_i=14m_p$ ($m_p:$ proton mass) are $\omega_{ci} = ZeB_0/m_i$, $\omega_{pi}={(Z^2e^2n_0/\epsilon_0 m_i)}^{1/2}$ and $\omega_{LH}= {({(\omega_{ce}\omega_{ci})}^{-1}+\omega_{pi}^{-2})}^{-1/2}$. The Alfv\'en speed is $v_A = B_0/{(\mu_0n_0m_i)}^{0.5} \approx 7.9 \times 10^5$ m/s. The ion acoustic speed is $c_s={((\gamma_eZk_BT_e + \gamma_ik_BT_i)/m_i)}^{1/2} \approx 2.2 \times 10^5$ m/s for $\gamma_e = 1$ and $\gamma_i = 3$. The speed $V_{fms}={(c_s^2 + v_A^2)}^{1/2} = 8.2 \times 10^5$ m/s determines the dispersion relation $\omega = V_{fms}k$ of the low-frequency modes. 

The simulation box covers with $x=\hat{x}/r_{ge}$ the interval $-2000 < x < 4000$. Wave numbers are multiplied with $r_{ge}$. Unless stated otherwise, times are given in units of $\omega_{lh}^{-1}$ and frequencies in units of $\omega_{lh}$. The ion density $n_{ion}$ is expressed in units of $n_0$. We examine the late times $T_0 \le t \le T_{max}$ with $T_0 = 461$ (190 ns). We resolve $T_{max}= 553$ (227 ns) by $1.4 \times 10^7$ time steps, which exceeds that in Ref. \cite{Dieckmann16} by the factor $100$.

\begin{figure*}
\includegraphics[width=15cm]{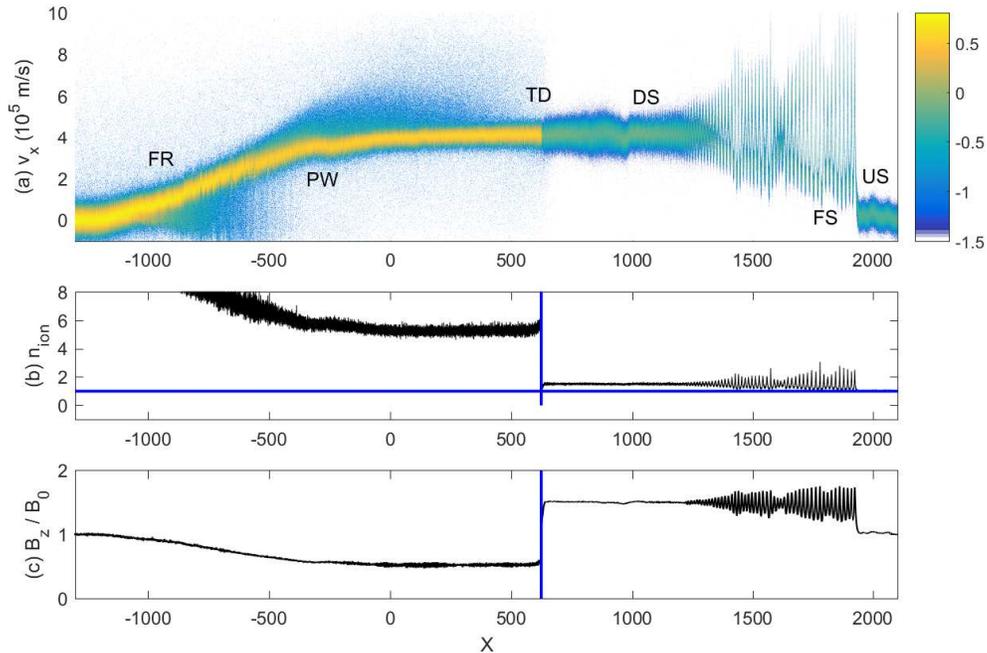}
\caption{Panel (a) shows the phase space density distribution of the ions on a 10-logarithmic scale. We recognize the fast rarefaction wave (FR), the precursor wave (PW), the tangential discontinuity (TD) and the fast magnetosonic shock (FS). The downstream region is indicated by DS and the upstream region by US. A weak ion beam is present ahead of the shock, which is not visible in the still frame. (b) shows the ion density $n_{ion}$. The blue lines denote $n_{ion}=1$ and $x=625$. (c) plots the $B_z$ component. The blue line denotes $x=625$. The time is $t=T_0$ (Multimedia view).}
\label{figure1}
\end{figure*}

Figure \ref{figure1} shows the ion phase space density, the ion density and the magnetic field at the time $T_0$.
The front of the rarefaction wave has reached in Fig. \ref{figure1}(a) the position $x\approx -1300$. The ion density and the amplitude of $B_z$ decrease and the ion speed increases with increasing $x$. This structure is a fast rarefaction wave. It expands up to $x\approx -400$ and ends in a precursor wave that is confined to the end of the rarefaction wave. The variation of the magnetic field amplitude and density across the precursor wave are in phase and it is a fast mode. It is spatially damped in the direction of larger $x$. The mean velocity, temperature and density of the ions remain approximately constant until $x=625$, where the ion density decreases, while the ion temperature and the amplitude of $B_z$ increase. This structure is a tangential discontinuity.  It is stable and long-lived. The ion distribution, density and the magnetic field remain unchanged in the interval $625 < x <1300$. The oscillations within $1300<x<1950$ correspond to the transition layer of a fast magnetosonic shock. We define the downstream region as the interval $625 < x < 1300$ that is enclosed by the shock and the tangential discontinuity.   

A dilute population of hot ions is found to the left of the discontinuity and it is confined by it. These ions are gradually accelerated up to a velocity modulus $\approx 5\times 10^5$ m/s. They are accelerated by their interaction with electrostatic fluctuations, which are strong due to the large electron temperature and plasma density \cite{Dieckmann04}. The denser population of hot ions in the interval $-1000<x<-500$ and $v_x \approx 0$ in Fig. \ref{figure1}(a) has also been observed at the front of unmagnetized rarefaction waves \cite{Sarri10} and is thus probably tied to ion acceleration by the spatially nonuniform electric field noise. The ions reach a peak speed of $2.5\times 10^6$ m/s (not shown). Some ions travel from the shock ahead of it. Their number density is too low to enforce a shock reformation.

\begin{figure}
\includegraphics[width=\columnwidth]{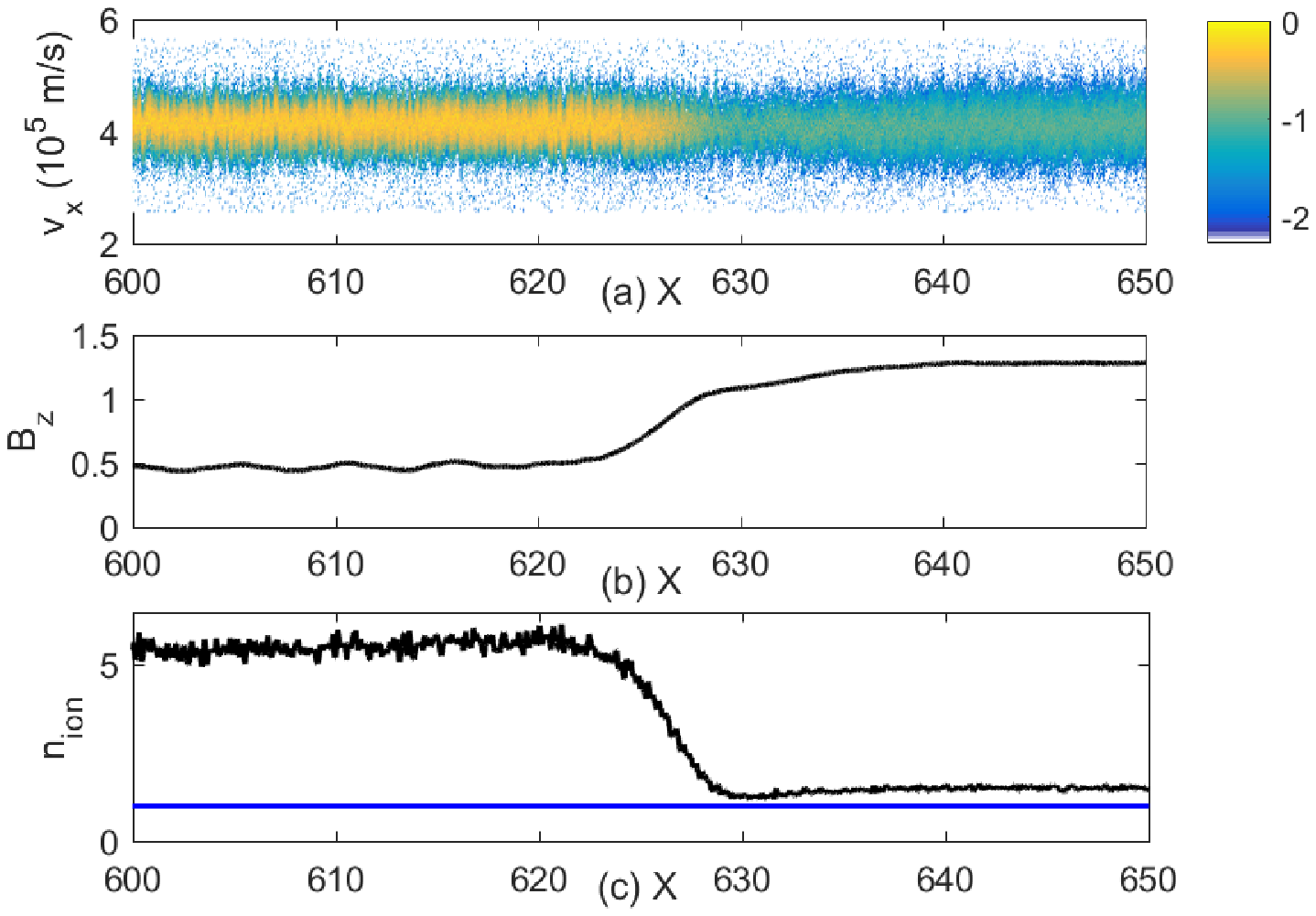}
\caption{Panel (a) shows the 10-logarithm of the ion phase space density normalized to its peak value close to the tangential discontinuity. (b) shows the distribution of $B_z$ and (c) that of $n_{ion}$ and the blue line corresponds to $n_{ion}=1$. The simulation time is $T_0$.}\label{figure2}
\end{figure}

Figure \ref{figure2} examines the ion- and magnetic field distributions close to the tangential discontinuity at $x\approx 625$.
Figure \ref{figure2}(a) shows that the ions move at the spatially uniform mean speed $v_b \approx 4.1 \times 10^5$ m/s or $v_b \approx V_{fms}/2$. The simulation frame equals the upstream frame and $v_b$ is thus the speed of the downstream plasma in the upstream frame. The ion phase space density, the value of $B_z$ and that of $n_{ion}$ change rapidly over 5$r_{ge}$ and reach their respective downstream values $B_z \approx 1.3$ T and $n_{ion}\approx 1.5$ at $x = 640$. The change in $B_z$ is sustained by an electron drift along $y$ and the electron temperature to the right of the discontinuity is about 100 eV below that to the left, which is about 3 keV (not shown). 

The density change at $x\approx 625$ yields a thermal pressure gradient force. The thermal pressure is $P_{th}(x) \approx n_e(x) k_B T_j$ and its change is $\Delta_P = P_{th}(x>625)-P_{th}(x<625)$. The gradient of the magnetic pressure $P_B = B_z^2(x)/2\mu_0$ yields a force that points in the opposite direction. Figure \ref{figure2} suggests changes in the electron density and magnetic field amplitude of 25$n_0$ and $B_0$, which gives $|P_B(x>625)-P_B(x<625)|\approx 0.6\Delta_P$. The moving magnetic field structure is exposed to the ram pressure of the upstream medium at $x\approx 1950$. The ram pressure $P_R = n_0 m_i v_b^2 \approx 0.5 \Delta_P$ balances the difference between both pressures at $x=625$. The pressure balance implies that the boundary is stationary in the downstream frame. There is no net ion flow across this tangential discontinuity since the Larmor radius of the energetic ions, which move with a few 100 km/s in the downstream frame, is only about 100$r_g$. 

\begin{figure}
\includegraphics[width=\columnwidth]{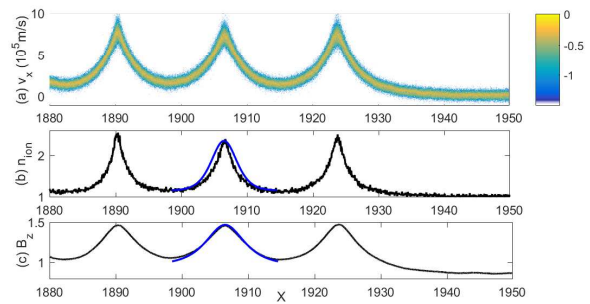}
\caption{The front of the fast magnetosonic shock. Panel (a) shows the 10-logarithmic ion phase space density distribution normalized to its peak value. (b) shows the ion density $n_{ion}$ and a fit of the function $sech^2(k_0\tilde{x})$ (blue curve). (c) shows the magnetic field $B_z$ and the fit $sech(k_0\tilde{x})$ (blue curve). We used $k_0 = 2\pi/16$ and $\tilde{x}=x-1906.5$.}\label{figure3}
\end{figure}

Figure \ref{figure3} shows the distributions of the ion phase space density, the ion density and the magnetic field at the front of the fast magnetosonic shock. The upstream ions are located in the interval $x>1940$.
Figure \ref{figure3}(a) reveals ion velocity oscillations with an amplitude $\approx v_b$. The distribution shows cusps at the maxima and the waves are not linear. This is confirmed by the non-sinusoidal oscillations of the ion density and the magnetic field distributions in Figs. \ref{figure3}(b,c). The magnetic field distribution is approximated well by a hyperbolic secant and the density follows approximately its square. The magnetic pressure $\propto B_z^2$ follows the thermal pressure $\propto n_{ion}$, which suggests that these waves are fast magnetosonic waves. 

The dispersion relation of the structures in the shock transition layer will determine the underlying wave modes. These are confined to the downstream plasma and hence we must evaluate their properties in the downstream frame. The ion density and magnetic field are $1.5n_{ion}$ and 1.3T in the downstream region, which give a lower-hybrid frequency $\omega_{lh}^* \approx 1.5 \omega_{lh}$. We select $x^* = x-v_bt^*-1920$ and $t^* = (t-T_0)$ as the transformation from the box frame into the downstream frame. 

\begin{figure}
\includegraphics[width=\columnwidth]{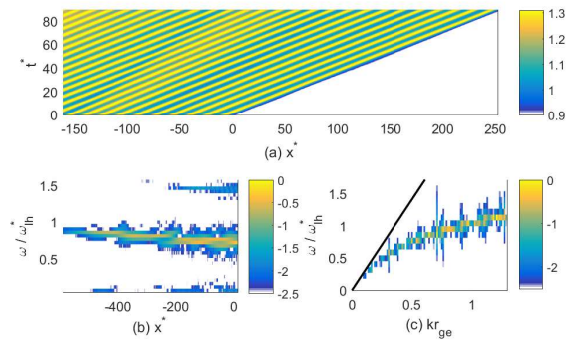}
\caption{Panel (a) shows the magnetic field $B_z(x^*,t^*)$ sampled in the downstream frame of reference. (b) shows the power spectrum of the Fourier transform of $B_z(x^*,t^*)$ over time. (c) compares the dispersion relation of the fast magnetosonic waves to the dispersion relation $\omega / k = V^*_{fms}$ of the fast magnetosonic mode in the low $k$ approximation.}\label{figure4}
\end{figure}

Figure \ref{figure4}(a) depicts $B_z(x^*,t^*)$ at the front of the expanding plasma. The speed of the wave front is $v_f \approx 8.5 \times 10^5$ m/s. This speed corresponds to the shock speed measured in the downstream frame $x^*$ and it is thus well below the fast magnetosonic speed $V^*_{fms} \approx 1.3 \times 10^6$ m/s in the downstream plasma. The speed of the wave front in the upstream medium is supersonic with $v_f + v_b \approx 1.5 V_{fms}$.
The power spectrum of $B_z(x^*,\omega)$ in Fig. \ref{figure4}(b) peaks below $\omega_{lh}^*$ and the wave frequency, at which the spectrum peaks, decreases with increasing $x^*$. A wave harmonic is observed close to $x^* \approx 0$, where the amplitude of $B_z$ is in the nonlinear regime (See Fig. \ref{figure3}(c)).

Figure \ref{figure4}(c) compares the dispersion relation of the fast magnetosonic mode $\omega / k = V^*_{fms}$ in the electromagnetic limit $k\rightarrow 0$ to the power spectrum of the noise, which we measured in a separate PIC simulation. That simulation modelled a spatially uniform plasma in a thermal equilibrium with the plasma parameters of the downstream region in Fig. \ref{figure1}(a). The noise distribution in PIC simulations peaks at values $(\omega,k)$, which correspond to eigenmodes of the system \cite{Dieckmann04}. The dispersion relation of the noise takes into account also the electrostatic component of the fast magnetosonic mode, which becomes important close to $\omega^*_{lh}$. The wave power in Fig. \ref{figure4}(b) peaks in the frequency band $0.75 < \omega / \omega^*_{lh} < 1$, where the phase speed of the fast magnetosonic wave is well below $V^*_{fms}$
and where the phase speed decreases with increasing $k$. The dispersion relation thus explains firstly why the wave front in Fig. \ref{figure4}(a) moves at the speed $v_f \approx 0.65 V^*_{fms}$ and, secondly, why the wave frequency decreases with increasing $x^*$ in Fig. \ref{figure4}(b). The steepening of the fast magnetosonic shock results in waves with a larger $k$ that fall behind the shock due to their lower phase speed.

In summary we have tracked with a 1D PIC simulation the expansion of a dense plasma into a magnetized ambient plasma over unprecedented time scales and we could observe the emergence of MHD structures. Their emergence was made possible by the low speed of the shock, which allowed it to dissipate the directed flow energy of the inflowing plasma without the need of reflecting many of its ions back upstream. We observed a fast rarefaction wave, which ended in a precursor wave, a tangential discontinuity and a fast magnetosonic shock. The shock involved large wave numbers, in which the fast magnetosonic wave branch is dispersive. We have shown for the first time that the dispersive nature of the fast magnetosonic wave branch transforms the fast magnetosonic shock into a train of non-linear oscillations, which gives rise to a broad shock transition layer. 

M. E. D. acknowledges financial support by a visiting fellowship of CRAL. The simulations were performed on resources provided by the Grand Equipement National de Calcul Intensif (GENCI) through grant x2016046960 and by the Swedish National Infrastructure for Computing (SNIC) at HPC2N (Ume\aa).


\begin{thebibliography}{}
\bibitem{Dolder60} K. Dolder, and R. Hide, Rev. Mod. Phys. \textbf{32} 770 (1960).
\bibitem{Borisov71} M. B. Borisov, S. G. Zaitsev, E. I. Chebotareva, and E. V. Lazareva, Fluid Dyn. \textbf{6} 501 (1971).
\bibitem{Brio88} M. Brio, and C. C. Wu, J. Comput. Phys. \textbf{75} 400 (1988).
\bibitem{Falle98} S. A. E. G. Falle, S. S. Komissarov, and P. Joarder, Mon. Not. R. Astron. Soc. \textbf{297} 265 (1998).
\bibitem{Balogh05} A. Balogh, S. J. Schwartz, S. D. Bale, M. A. Balikhin, D. Burgess, T. S. Horbury, V. V. Krasnoselskikh, H. Kucharek, B. Lembege, E. A. Lucek, E. Mobius, M. Scholer, M. F. Thomsen, and S. N. Walker, Space Sci. Rev. \textbf{118}, 155 (2005).
\bibitem{Stasiewicz03} K. Stasiewicz, M. Longmore, S. Buchert, P. K. Shukla, B. Lavraud, and J. Pickett, Geophys. Res. Lett. \textbf{30} 2241 (2003).
\bibitem{Guerolt17} R. Guerolt, Y. Ohsawa, and N. J. Fisch, Phys. Rev. Lett. \textbf{118} 125101 (2017).
\bibitem{Eastwood05} J. P. Eastwood, E. A. Lucek, C. Mazelle, K. Meziane, Y. Narita, J. Pickett, and R. A. Treumann, Space Sci. Rev. \textbf{118} 41 (2005).
\bibitem{Chapman05} S. C. Chapman, R. E. Lee, and R. O. Dendy, Space Sci. Rev. \textbf{121} 5 (2005).
\bibitem{Burgess07} D. Burgess, and M. Scholer, Phys. Plasmas \textbf{14} 012108 (2007).
\bibitem{Marcowith16} A. Marcowith, A. Bret, A. Bykov, M. E. Dieckman, L. O. Drury, B. Lembege, M. Lemoine, G. Morlino, G. Murphy, G. Pelletier, I. Plotnikov, B. Reville, M. Riquelme, L. Sironi, and A. S. Novo, Rep. Prog. Phys. \textbf{79} 046901 (2016).
\bibitem{Sundberg17} T. Sundberg, D. Burgess, M. Scholer, A. Masters, and A. H. Sulaiman, Astrophys. J. \textbf{836} L4 (2017).
\bibitem{Arber15} T. D. Arber, K. Bennett, C. S. Brady, A. Lawrence-Douglas, M. G. Ramsay, N. J. Sircombe, P. Gillies, R. G. Evans, H. Schmitz, A. R. Bell, and C. P. Ridgers, Plasma Phys. Controll. Fusion \textbf{57} 113001 (2015).
\bibitem{Dieckmann16} M. E. Dieckmann, G. Sarri, D. Doria, A. Ynnerman, and M. Borghesi, Phys. Plasmas \textbf{23} 062111 (2016).
\bibitem{Myong97} R. S. Myong, and P. L. Roe, J. Plasma Phys. \textbf{58} 521 (1997).
\bibitem{Dieckmann04} M. E. Dieckmann, A. Ynnerman, S. C. Chapman, G. Rowlands, and N. Andersson, Phys. Scripta \textbf{69} 456 (2004).
\bibitem{Sarri10} G. Sarri, M. E. Dieckmann, I. Kourakis, and M. Borghesi, Phys. Plasmas \textbf{17} 082305 (2010).
\end{thebibliography}
\end{document}